\documentclass[12pt]{article}
\usepackage{amssymb}
\renewcommand{\theequation}{\arabic{section}.\arabic{equation}}

\begin{document}



\def\a{\alpha}
\def\b{\beta}
\def\d{\delta}
\def\e{\epsilon}
\def\g{\gamma}
\def\h{\mathfrak{h}}
\def\k{\kappa}
\def\l{\lambda}
\def\o{\omega}
\def\p{\wp}
\def\r{\rho}
\def\t{\tau}
\def\s{\sigma}
\def\z{\zeta}
\def\x{\xi}
\def\V={{{\bf\rm{V}}}}
 \def\A{{\cal{A}}}
 \def\B{{\cal{B}}}
 \def\C{{\cal{C}}}
 \def\D{{\cal{D}}}
\def\G{\Gamma}
\def\K{{\cal{K}}}
\def\O{\Omega}
\def\R{\bar{R}}
\def\T{{\cal{T}}}
\def\L{\Lambda}
\def\f{E_{\tau,\eta}(sl_2)}
\def\E{E_{\tau,\eta}(sl_n)}
\def\Zb{\mathbb{Z}}
\def\Cb{\mathbb{C}}

\def\R{\overline{R}}

\def\beq{\begin{equation}}
\def\eeq{\end{equation}}
\def\bea{\begin{eqnarray}}
\def\eea{\end{eqnarray}}
\def\ba{\begin{array}}
\def\ea{\end{array}}
\def\no{\nonumber}
\def\le{\langle}
\def\re{\rangle}
\def\lt{\left}
\def\rt{\right}

\newtheorem{Theorem}{Theorem}
\newtheorem{Definition}{Definition}
\newtheorem{Proposition}{Proposition}
\newtheorem{Lemma}{Lemma}
\newtheorem{Corollary}{Corollary}
\newcommand{\proof}[1]{{\bf Proof. }
        #1\begin{flushright}$\Box$\end{flushright}}

\baselineskip=20pt

\newfont{\elevenmib}{cmmib10 scaled\magstep1}
\newcommand{\preprint}{
   \begin{flushleft}
   \end{flushleft}\vspace{-1.3cm}
   \begin{flushright}\normalsize
   \end{flushright}}
\newcommand{\Title}[1]{{\baselineskip=26pt
   \begin{center} \Large \bf #1 \\ \ \\ \end{center}}}
\newcommand{\Author}{\begin{center}
   \large \bf
Kun Hao${}^{a}$,~Junpeng Cao${}^{b,c}$,~Guang-Liang Li${}^{d}$,~Wen-Li Yang${}^{a,e}\footnote{Corresponding author:
wlyang@nwu.edu.cn}$,
 ~ Kangjie Shi${}^a$ and~Yupeng Wang${}^{b,c}\footnote{Corresponding author: yupeng@iphy.ac.cn}$
 \end{center}}
\newcommand{\Address}{\begin{center}

     ${}^a$Institute of Modern Physics, Northwest University,
     Xian 710069, China\\
     ${}^b$Beijing National Laboratory for Condensed Matter
           Physics, Institute of Physics, Chinese Academy of Sciences, Beijing
           100190, China\\
     ${}^c$Collaborative Innovation Center of Quantum Matter, Beijing,
     China\\
     ${}^d$Department of Applied Physics, Xian Jiaotong University, Xian 710049, China\\
     ${}^e$Beijing Center for Mathematics and Information Interdisciplinary Sciences, Beijing, 100048,  China

   \end{center}}
\newcommand{\Accepted}[1]{\begin{center}
   {\large \sf #1}\\ \vspace{1mm}{\small \sf Accepted for Publication}
   \end{center}}

\preprint
\thispagestyle{empty}
\bigskip\bigskip\bigskip

\Title{Exact solution  of the Izergin-Korepin model with general non-diagonal boundary terms   }
\Author

\Address
\vspace{1cm}

\begin{abstract}
The Izergin-Korepin model with general non-diagonal boundary terms, a typical integrable model beyond
 A-type and without $U(1)$-symmetry, is studied via the off-diagonal Bethe ansatz method.
Based on some intrinsic properties of the $R$-matrix and the $K$-matrices, certain operator product identities of
the transfer matrix are obtained at some special points of the spectral parameter. These identities
and the asymptotic behaviors of the transfer matrix together allow us to construct  the  inhomogeneous
$T-Q$  relation and  the associated Bethe ansatz equations. In the diagonal boundary limit,
the reduced results coincide exactly with those obtained via other methods.

\vspace{1truecm} \noindent {\it PACS:} 75.10.Pq, 03.65.Vf, 71.10.Pm

\end{abstract}
\newpage
\section{Introduction}
\label{intro} \setcounter{equation}{0}
Exactly solvable models (or quantum integrable systems) \cite{Bax82} have
provided important benchmarks for the non-perturbative analysis of quantum systems
appearing in  string and super-symmetric Yang-Mills (SYM) theories \cite{Dol03}
(see also \cite{Bei12} and references therein), low-dimensional condensed matter
physics \cite{Zvy05,Gua13} and statistical physics \cite{Gie05-1,Sir09}. However,
it has been well known for many years that there exists a quite usual class of integrable
models, which do not possess $U(1)$-symmetry and thus make the conventional Bethe ansatz methods almost inapplicable.
Some famous examples are the closed XYZ chain with odd number of sites
\cite{Tak79}, the anisotropic spin
torus \cite{Yun95}, the quantum spin chains with non-diagonal
boundary fields \cite{Nep04,Cao03} and their multi-component generalizations \cite{Yan04,Gal05,Yan05,Cao13}.
The broken $U(1)$-symmetry in those models
leads to the absence of an obvious reference state, which is crucial for the usage of
the conventional Bethe ansatz (BA) methods \cite{Bet31,Alc87,bax1,Bax82,Skl78,Tak79,Skl88}.  There have been
numerous efforts \cite{Nep04,Cao03,Gie05,Yan04-1,Gie05-1,Doi06,Baj06,Bas07,Yan06,Gal08,Fra08,Nic12,Cao1}
to approach the exact solutions \footnote{It is also interesting that the eigenstates can been classified by the representation of
the so-called q-Onsager algebra \cite{Bac06,Bac13} and that the half-infinite XXZ spin chain with a triangular boundary has been
studied by q-vertex operator method \cite{Bac14}. } to this class of models over the last 20 years.

Very recently, based on the intrinsic
properties of the $R$-matrix and the $K$-matrices for quantum integrable models, a systematic method for
identifying the spectrum of integrable models without $U(1)$-symmetry, i.e., the
off-diagonal Bethe ansatz (ODBA) method was proposed
in \cite{Cao1} for the models associated with $su(2)$ algebra. Subsequently,
 the nested-version of ODBA for
the models associated with $su(n)$ algebra was developed in \cite{Cao13}.
Several long-standing models were then
diagonalized exactly \cite{Cao1,Cao13,Li14,Zha13} by the ODBA method \footnote{The completeness of the solutions for
the (an-)isotropic spin-$\frac{1}{2}$ chain has been checked numerically within the Bethe ansatz method \cite{Nep13,Jia13},
and holds by construction in the  quantum separation variables method (SOV) \cite{Fal13,Kit14} or  the q-Onsager algebra method \cite{Bas07}. An
expression for the corresponding eigenvectors for the XXX open chain in the framework of algebraic Bethe ansatz was also proposed
recently in \cite{Bel13-1}. }. Despite those progresses, an important issue, i.e., whether the ODBA method
can be applied to other multi-component integrable models defined beyond the $su(n)$ algebra, is still open.

In this paper, we study the Izergin-Korepin (IK) model \cite{Ize81} with generic integrable boundaries. This model has played a fundamental role in
quantum integrable models associated with algebras beyond $su(n)$ (or non A-type models).
It was introduced as a quantum integrable model related
to the Dodd-Bullough-Mikhailov or Jiber-Mikhailov-Shabat model \cite{Dod77,Zhi79}, one of two integrable relativistic models
containing one scalar field (the other is sine-Gordon model). The  $R$-matrix of the model corresponds to  the simplest twisted affine
algebra $A^{(2)}_2$. The IK model with open boundary condition is related to the loop models \cite{Yun95-1} and self-avoiding walks at a boundary \cite{Bat95}.
The IK model with $U(1)$-symmetry, i.e. with periodic boundary condition or  with diagonal boundaries have been extensively
studied \cite{Res83,Tar88,Kim94,Mat95,Mez92,Yun95-1,Fan97,Hou99,Lim99,Yan01,Nep02,Li03}.
Even the most general integrable boundary condition (corresponding to the non-diagonal reflection matrix) has been
known almost for 20 years \cite{Kim94,Lim99,Nep02},
its exact solution is still
missing. The purpose of this paper is to propose the spectrum of this model using  the ODBA method.

The paper is organized as follows.  Section 2 serves as an
introduction to the model and our notations. In Section 3, we derive certain operator
product identities for the transfer matrix of the model with general non-diagonal
boundary terms by using some intrinsic properties of the $R$-matrix and $K$-matrices.
The asymptotic behaviors of the transfer matrix are also obtained. Section 4 is devoted to the construction of the inhomogeneous $T-Q$ relation and the corresponding Bethe ansatz equations.
Section 5 is attributed to the reduction
to case of diagonal boundaries. It is found that the reduced results coincide exactly with those obtained
by other Bethe ansatz methods. In section 6, we summarize our results and give some discussions.
Some detailed technical proof is given in Appendix A.


\section{ Transfer matrix}
\label{XXZ} \setcounter{equation}{0}

Throughout, ${\rm\bf V}$ denotes a three-dimensional linear space and let $\{|i\rangle|i=1,2,3\}$ be an orthonormal basis of it.
We shall adopt the standard notations: for any matrix $A\in {\rm End}({\rm\bf V})$, $A_j$ is an
embedding operator in the tensor space ${\rm\bf V}\otimes
{\rm\bf V}\otimes\cdots$, which acts as $A$ on the $j$-th space and as
identity on the other factor spaces; For $B\in {\rm End}({\rm\bf V}\otimes {\rm\bf V})$, $B_{ij}$ is an embedding
operator of $B$ in the tensor space, which acts as identity
on the factor spaces except for the $i$-th and $j$-th ones.

The $R$-matrix $R(u)\in {\rm End}({\rm\bf V}\otimes {\rm\bf V})$ of
the IK model is given by \cite{Ize81} \bea
 R_{12}(u)=\left(\begin{array}{r|r|r}{\begin{array}{rrr}c(u)&&\\&b(u)&\\&&d(u)\end{array}}
           &{\begin{array}{lll}&&\\e(u)&{\,\,\,\,\,\,\,\,\,\,}&{\,\,\,\,\,\,\,\,\,\,}\\{\,\,\,\,\,\,}&g(u)&{\,\,\,\,\,\,}\end{array}}
           &{\begin{array}{lll}&&\\&&\\f(u)&{\,\,\,\,\,\,\,\,\,\,}&{\,\,\,\,\,\,\,\,\,\,}\end{array}}\\[12pt]
 \hline {\begin{array}{rrr}&\bar{e}(u)&\\&&\bar{g}(u)\\&&\end{array}}&
           {\begin{array}{ccc}b(u)&&\\&a(u)&\\&&b(u)\end{array}}
           &{\begin{array}{lll}&&\\g(u)&{\,\,\,\,\,\,\,\,\,\,}&{\,\,\,\,\,\,\,\,\,\,}\\&e(u)&{\,\,\,\,\,\,\,\,\,\,}\end{array}}\\[12pt]
 \hline {\begin{array}{ccc}&&\bar{f}(u)\\&&\\&&\end{array}}
           &{\begin{array}{ccc}&\bar{g}(u)&\\&&\bar{e}(u)\\&&\end{array}}
           &{\begin{array}{ccc}d(u)&&\\&b(u)&\\&&c(u)\end{array}} \end{array}\right),\label{R-matrix}
\eea \noindent where the matrix elements are \bea
&&a(u)=\sinh(u\hspace{-0.04truecm}-\hspace{-0.04truecm}3\eta)
\hspace{-0.04truecm}-\hspace{-0.04truecm}\sinh
5\eta\hspace{-0.04truecm}+\hspace{-0.04truecm} \sinh
3\eta\hspace{-0.04truecm}+\hspace{-0.04truecm}\sinh\eta,\,\,
b(u)=\sinh(u\hspace{-0.04truecm}-\hspace{-0.04truecm}3\eta)
\hspace{-0.04truecm}+\hspace{-0.04truecm}\sin3\eta,\no\\[6pt]
&&c(u)=\sinh(u-5\eta)+\sinh\eta,\quad d(u)=\sinh(u-\eta)+\sinh\eta,\no\\[6pt]
&&e(u)=-2e^{-\frac{u}{2}}\sinh2\eta\cosh(\frac{u}{2}-3\eta),\quad \bar{e}(u)=-2e^{\frac{u}{2}}\sinh2\eta\cosh(\frac{u}{2}-3\eta),\no\\[6pt]
&&f(u)=-2 e^{-u+2\eta}\sinh\eta\sinh2\eta-e^{-\eta}\sinh4\eta,\no\\[6pt]
&&\bar{f}(u)=2 e^{u-2\eta}\sinh\eta\sinh2\eta-e^{\eta}\sinh4\eta,\no\\[6pt]
&&g(u)=2e^{-\frac{u}{2}+2\eta}\sinh\frac{u}{2}\sinh 2\eta,\quad \bar{g}(u)=-2e^{\frac{u}{2}-2\eta}\sinh\frac{u}{2}\sinh 2\eta.
\label{R-element-2}
\eea The $R$-matrix satisfies the quantum Yang-Baxter equation (QYBE)
\bea
R_{12}(u_1-u_2)R_{13}(u_1-u_3)R_{23}(u_2-u_3)
=R_{23}(u_2-u_3)R_{13}(u_1-u_3)R_{12}(u_1-u_2),\label{QYB}\eea
and possesses the following   properties,
\bea &&\hspace{-1.5cm}\mbox{ Initial
condition}:\,R_{12}(0)= (\sinh\eta-\sinh5\eta)P_{12},\label{Int-R}\\
&&\hspace{-1.5cm}\mbox{ Unitarity
relation}:\,R_{12}(u)R_{21}(-u)= \rho_1(u)\,\times {\rm id},\label{Unitarity}\\
&&\hspace{-1.5cm}\mbox{ Crossing
relation}:\,R_{12}(u)=V_1R_{12}^{t_2}(-u+6\eta+i\pi)V^{-1}_1,
\label{crosing}\\
&&\hspace{-1.5cm}\mbox{ PT-symmetry}:\,R_{21}(u)=R^{t_1\,t_2}_{12}(u),\label{PT}\\
&&\hspace{-1.5cm}\mbox{ Periodicity}: \,
\qquad R_{12}(u+2i\pi)=R_{12}(u).\label{Periodic}
\eea
Here $R_{21}(u)=P_{12}R_{12}(u)P_{12}$ with $P_{12}$ being
the usual permutation operator and $t_i$ denotes transposition in
the $i$-th space. The function $\rho_1(u)$ and the crossing matrix $V$ are given by
\bea
\rho_1(u)&=&-4\sinh(\frac{u}{2}-2\eta)\sinh(\frac{u}{2}+2\eta)\cosh(\frac{u}{2}-3\eta)\cosh(\frac{u}{2}+3\eta),\label{p-1-function}\\[6pt]
V&=&\left(\begin{array}{ccc}&&-e^{-\eta}\\&1&\\-e^{\eta}\end{array}\right),\quad V^2=1.\label{V-matrix}
\eea

The unitarity property (\ref{Unitarity}) and crossing relation (\ref{crosing}) of the $R$-matrix and expressions (\ref{p-1-function})-(\ref{V-matrix})
of  the  function $\rho_1(u)$ and the $V$-matrix  imply that the $R$-matrix satisfies the crossing-unitarity relation
\bea
&&R_{12}^{t_1}(u)\,{\cal M}_1\,R_{21}^{t_1}(-u+12\eta)\,{\cal M}_1^{-1}=\rho_2(u)\,\times {\rm id},\label{crossing-unitarity} \\[6pt]
&& {\cal M}=V^{t}\,V=\left(\begin{array}{ccc}e^{2\eta}&&\\&1&\\&&e^{-2\eta}\end{array}\right),\label{M-matrix}\\[6pt]
&&\rho_2(u)=-4\cosh(\frac{u}{2}-5\eta)\cosh(\frac{u}{2}-\eta)\sinh\frac{u}{2}\sinh(\frac{u}{2}-6\eta).\label{p-2-function}
\eea It is easily to check that the $R$-matrix also satisfies the following relation
\bea
{\cal M}_1\,{\cal M}_2\,R_{12}(u)\,{\cal M}_1^{-1}\,{\cal M}_2^{-1}=R_{12}(u),\quad R_{12}(u)\,R_{21}(v)=R_{21}(v)\,R_{12}(u).\label{Invariant}
\eea

Let us introduce the ``row-to-row"  (or one-row ) monodromy matrices
$T(u)$ and $\hat{T}(u)$, which are  $3\times 3$ matrices with operator-valued elements acting on ${\rm\bf V}^{\otimes N}$,
\bea
T_0(u)&=&R_{0N}(u-\theta_N)R_{0\,N-1}(u-\theta_{N-1})\cdots
R_{01}(u-\theta_1),\label{Mon-V-1}\\
\hat{T}_0(u)&=&R_{10}(u+\theta_1)R_{20}(u+\theta_{2})\cdots
R_{N0}(u+\theta_N).\label{Mon-V-2}
\eea Here $\{\theta_j|j=1,\cdots,N\}$ are
arbitrary free complex parameters which are usually called as
inhomogeneous parameters. The transfer matrix can be constructed as follows \cite{Skl88,Mez91}.
Let us introduce further a pair of $K$-matrices $K^-(u)$ and $K^+(u)$. The
former satisfies the reflection equation (RE)
 \bea &&R_{12}(u_1-u_2)K^-_1(u_1)R_{21}(u_1+u_2)K^-_2(u_2)\no\\
 &&~~~~~~=
K^-_2(u_2)R_{12}(u_1+u_2)K^-_1(u_1)R_{21}(u_1-u_2),\label{RE-V}\eea
and the latter  satisfies the dual RE \bea
&&R_{12}(u_2-u_1)K^+_1(u_1){\cal M}_1^{-1}R_{21}(-u_1-u_2+12\eta){\cal M}_1K^+_2(u_2)\no\\[6pt]
&&~~~~~~= K^+_2(u_2){\cal M}_2^{-1}R_{12}(-u_1-u_2+12\eta){\cal M}_2K^+_1(u_1)R_{21}(u_2-u_1).
\label{DRE-V}\eea For open boundaries,  one needs to
consider  the
 double-row monodromy matrix $\mathbb{T}(u)$
\bea
  \mathbb{T}(u)=T(u)K^-(u)\hat{T}(u).
  \label{Mon-V-0}
\eea The double-row transfer matrix $t(u)$ is thus given by
\bea
t(u)=tr(K^+(u)\mathbb{T}(u)).\label{trans}\eea The QYBE (\ref{QYB})
and (dual) REs (\ref{RE-V}) and (\ref{DRE-V}) lead to the fact that
the transfer matrices with different spectral parameters commute
with each other \cite{Skl88}: $[t(u),t(v)]=0$. Therefore $t(u)$
serves as the generating functional of the conserved quantities of
the corresponding system.

In this paper we consider the generic non-diagonal $K$-matrix
$K^-(u)$ found in \cite{Kim94} and was classified as type II
\footnote{The generalization to the  non-diagonal $K$-matrices of
type I in \cite{Lim99,Nep02} is straightforward.}in
\cite{Lim99,Nep02} \bea
K^-(u)=\left(\begin{array}{ccc}1+2e^{-u-\e}\sinh\eta&0&2e^{-\e+\sigma}\sinh u\\
0&1-2e^{-\e}\sinh(u-\eta)&0\\
2e^{-\e-\sigma}\sinh u&0&1+2e^{u-\e}\sinh\eta\end{array}\right),
\label{K-matrix-1} \eea \noindent and non-diagonal $K^+(u)$  given
by \bea K^+(u)={\cal
M}K^-(-u+6\eta+i\pi)\left|_{(\e,\sigma)\rightarrow
(\e',\sigma')}\right..\label{K-matrix-2} \eea Besides the crossing
parameter $\eta$, the corresponding transfer matrix $t(u)$ given by
(\ref{trans}) has four other free parameters
$\{\e,\,\sigma,\,\e',\,\sigma'\}$  describing the boundary fields.
The Hamiltonian of the Izergin-Korepin model with general
non-diagonal boundary terms specified by the $K$-matrices given by
(\ref{K-matrix-1}) and (\ref{K-matrix-2}) then is given in terms of
the transfer matrix by
\begin{eqnarray}
&&H=
\left.\frac{\partial \ln t(u)}{\partial
u}\right|_{u=0,\{\theta_j=0\}} \nonumber \\[6pt]
&&\quad = \sum_{j=1}^{N-1}\frac {2P_{jj+1}R^\prime_{jj+1}(0)}{\sinh \eta-
\sinh 5\eta} + \frac{tr {K^+}^\prime(0)}{tr K^+(0)} +\frac {2 tr_0
[K_{0}^+(0) P_{N0}R^\prime_{N0}(0)]}{ (\sinh \eta- \sinh 5\eta) tr K^+(0)}
+ \frac{{K_{1}^-}^\prime(0)}{1+2e^{-\epsilon} \sinh \eta } \nonumber
\\[6pt]
&& \quad =\frac 2{\sinh \eta- \sinh 5\eta}\sum_{j=1}^{N-1}\left\{
\cosh5\eta (E_{j}^{11}E_{j+1}^{11}+E_{j}^{33}E_{j+1}^{33})
\right. \nonumber \\[6pt]
&&\qquad \left.+\sinh2\eta(\sinh3\eta-\cosh3\eta)
(E_{j}^{11}E_{j+1}^{22}+E_{j}^{22}E_{j+1}^{33})
\right. \nonumber \\[6pt]
&&\qquad \left. +\sinh2\eta(\sinh3\eta+\cosh3\eta)
(E_{j}^{22}E_{j+1}^{11}+E_{j}^{33}E_{j+1}^{22})
\right. \nonumber \\[6pt]
&&\qquad \left.+2\sinh\eta \sinh2\eta (e^{-2\eta}
E_{j}^{11}E_{j+1}^{33}+e^{2\eta} E_{j}^{33}E_{j+1}^{11})+\cosh\eta
(E_{j}^{13}E_{j+1}^{31}+ E_{j}^{31}E_{j+1}^{13})
\right. \nonumber \\[6pt]
&&\qquad \left. +\cosh3\eta
(E_{j}^{12}E_{j+1}^{21}+E_{j}^{21}E_{j+1}^{12}
+E_{j}^{22}E_{j+1}^{22}+E_{j}^{23}E_{j+1}^{32}
+E_{j}^{32}E_{j+1}^{23}) \right. \nonumber \\[6pt]
&&\qquad \left.  -e^{-2\eta}\sinh2\eta
(E_{j}^{12}E_{j+1}^{32}+E_{j}^{21}E_{j+1}^{23}) +
e^{2\eta}\sinh2\eta (E_{j}^{23}E_{j+1}^{21}+E_{j}^{32}E_{j+1}^{12})
\right\} \nonumber \\[6pt]
&& \qquad -\frac {2e^{-\epsilon}}{1+2e^{-\epsilon} \sinh \eta}\left[
\sinh\eta (E_{1}^{11}-E_{1}^{33})+\cosh\eta E_{1}^{22} -e^{\sigma}
E_{1}^{13}-e^{-\sigma} E_{1}^{31} \right] \nonumber \\[6pt]
&& \qquad + 2 \left[(\sinh\eta-\sinh 5 \eta) (2\cosh 2\eta
-4e^{-\epsilon'} \sinh \eta\cosh 4\eta +1+ 2e^{-\epsilon'} \sinh
5\eta)\right]^{-1} \nonumber \\[6pt]
&& \qquad \times \left\{
\left[(e^{2\eta}-2e^{-4\eta-\epsilon'}\sinh\eta)\cosh5\eta+ (1+
2e^{-\epsilon'} \sinh 5\eta)
\right.\right. \nonumber \\[6pt]
&& \qquad \left.\left. \times \sinh 2\eta(\sinh 3\eta- \cosh 3 \eta)
+
2(e^{-4\eta}-2e^{2\eta-\epsilon'}\sinh\eta)\sinh\eta\sinh2\eta\right]E_{N}^{11}
\right. \nonumber \\[6pt]
&& \qquad \left. +
\left[(e^{2\eta}-2e^{-4\eta-\epsilon'}\sinh\eta)\sinh 2\eta(\sinh
3\eta+ \cosh 3 \eta) + (1+ 2e^{-\epsilon'} \sinh 5\eta)
\right.\right. \nonumber \\[6pt]
&& \qquad \left.\left. \times \cosh 3\eta +
2(e^{-2\eta}-2e^{4\eta-\epsilon'}\sinh\eta)\sinh 2\eta(\sinh 3\eta- \cosh 3
\eta) \right]E_{N}^{22}
\right. \nonumber \\[6pt]
&& \qquad \left. + \left[2(e^{4\eta}-2e^{-2\eta-\epsilon'}\sinh\eta)
\sinh \eta\sinh 2\eta + (1+ 2e^{-\epsilon'} \sinh 5\eta)
\right.\right. \nonumber \\[6pt]
&& \qquad \left.\left. \times \sinh 2\eta(\sinh 3\eta+ \cosh 3 \eta)
+
(e^{-2\eta}-2e^{4\eta-\epsilon'}\sinh\eta)\cosh5\eta\right]E_{N}^{33}
\right. \nonumber \\[6pt]
&& \qquad \left. - 2e^{-\epsilon'}\sinh6\eta\cosh\eta[
e^{2\eta+\sigma'}E_{N}^{13}+e^{-2\eta-\sigma'}E_{N}^{31}]
\right\} \nonumber \\[6pt]
&& \qquad +\frac{2e^{-\epsilon'}(2 \sinh \eta\sinh 4\eta- \cosh 5
\eta)} {2\cosh 2\eta -4e^{-\epsilon'} \sinh \eta\cosh 4\eta +1+
2e^{-\epsilon'} \sinh 5\eta}.
\end{eqnarray}
where $E_{j}^{\mu \nu}$ is the Weyl matrix or the Hubbard operator
\begin{eqnarray}
E^{\mu \nu}=|\mu \rangle \langle \nu |,\no
\end{eqnarray} and
\bea
R^\prime_{ij}(u)=\frac{\partial}{\partial u} R^\prime_{ij}(u),\quad
{K^{\pm}}^\prime_j(u)=\frac{\partial}{\partial u}K^{\pm}_j(u).
\eea


\section{ Operator identities of the transfer matrix}
\label{T-QR} \setcounter{equation}{0}

Following \cite{Cao13} we apply the fusion technique to study the the present model. In this
case, we need to use the fusion techniques both  for $R$-matrices
\cite{Kar79} and for $K$-matrices \cite{Mez92,Zho96}. Similar as that in \cite{Cao13}, the fusion
procedure will lead to the desired operator identities to determine the spectrum of the
transfer matrix $t(u)$ given by (\ref{trans}). For this purpose, let us introduce the following
vectors in the tensor space ${\rm\bf V}\otimes {\rm\bf V}$
\bea
|\Phi_0\rangle&=&\frac{1}{\sqrt{2\cosh 2\eta+1}}(e^{-\eta}|1,3\rangle-|2,2\rangle+e^{\eta}|3,1\rangle),\label{Vector-0}\\[6pt]
|\Phi_1\rangle&=&\frac{1}{\sqrt{2\cosh 2\eta}}(e^{-\eta}|1,2\rangle-e^{\eta}|2,1\rangle),\label{Vector-1}\\[6pt]
|\Phi_2\rangle&=&\frac{1}{\sqrt{2\cosh 2\eta}}(|1,3\rangle -2\sinh\eta|2,2\rangle -|3,1\rangle),\label{Vector-2}\\[6pt]
|\Phi_3\rangle&=&\frac{1}{\sqrt{2\cosh 2\eta}}(e^{-\eta}|2,3\rangle-e^{\eta}|3,2\rangle),\label{Vector-3}
\eea and the associated projectors \footnote{In contrast to most of rational models,  here $P^{(i)}_{12}\neq P^{(i)}_{21}$.}
\bea
P_{12}^{(1)}=|\Phi_0\rangle\langle\Phi_0|,\quad P_{12}^{(3)}=\sum_{i=1}^3|\Phi_i\rangle\langle\Phi_i|.\label{Projector}
\eea Direct calculation shows that the $R$-matrix given by (\ref{R-matrix}) at some degenerate points is proportional to the projectors,
\bea
R_{12}(6\eta+i\pi)=P^{(1)}_{12}\times S^{(1)}_{12},\quad R_{12}(4\eta)=P_{12}^{(3)}\times S^{(3)}_{12},\label{Fusion}
\eea where $S^{(i)}_{12}$ are some non-degenerate matrices $\in {\rm End}({\rm\bf V} \otimes {\rm \bf V})$. After some calculations, we find
\bea
&&P^{(1)}_{12} R_{23}(u)\,R_{13}(u+6\eta+i\pi)\,P^{(1)}_{12}=\rho_1(u)P^{(1)}_{12}\otimes {\rm id}={\rm Det}_q(R(u))P^{(1)}_{12}\otimes {\rm id},\label{Fused-R-1}\\
&&P^{(1)}_{12} R_{31}(u)\,R_{32}(u+6\eta+i\pi)\,P^{(1)}_{12}=\rho_1(u)P^{(1)}_{12}\otimes {\rm id},\label{Fused-R-2}\\[6pt]
&&P^{(3)}_{12} R_{23}(u)\,R_{13}(u+4\eta)\,P^{(3)}_{12}=\rho_3(u)\,R_{13}(u+2\eta+i\pi),\label{Fused-R-3}\\[6pt]
&&P^{(3)}_{12} R_{31}(u)\,R_{32}(u+4\eta)\,P^{(3)}_{12}=\rho_3(u)\,R_{31}(u+2\eta+i\pi),\label{Fused-R-4}
\eea where the function $\rho_3(u)$ is
\bea
\rho_3(u)=-2\sinh(\frac{u}{2}+2\eta)\cosh(\frac{u}{2}-3\eta).\label{p-3-function}
\eea
Noting that the rank of the projector $P^{(1)}_{12}$ is one, the quantum determinant \cite{Kul79} ${\rm Det}_q(R(u))$ is defined by
\bea
{\rm Det}_q(R(u))=tr_{12}\lt\{P^{(1)}_{12} R_{23}(u)\,R_{13}(u+6\eta+i\pi)\,P^{(1)}_{12}\rt\},\label{Q-Det-1}
\eea which plays the role of the generating function of the centers of the associated quantum algebras \cite{Cha94}. In our particular case
the corresponding   quantum determinant is proportional to the identity operator,
\bea
{\rm Det}_q(R(u))=\rho_1(u)\, {\rm id}.\label{Q-Det-2}
\eea
The above relations (\ref{Fused-R-1})-(\ref{Q-Det-2}) imply that the associated one-row monodromy matrices satisfy the following relations
\bea
&&P^{(1)}_{12} T_2(u)\,T_1(u+6\eta+i\pi)P^{(1)}_{12}=\prod_{l=1}^N \rho_1(u-\theta_l)\,P^{(1)}_{12}\otimes {\rm id},\label{Fused-Mon-matrix-1}\\
&&P^{(1)}_{12} \hat{T}_1(u)\,\hat{T}_2(u+6\eta+i\pi)P^{(1)}_{12}=\prod_{l=1}^N \rho_1(u+\theta_l)\,P^{(1)}_{12}\otimes {\rm id},\label{Fused-Mon-matrix-2}\\
&&P^{(3)}_{12} T_2(u)\,T_1(u+4\eta)P^{(1)}_{12}=\prod_{l=1}^N \rho_3(u-\theta_l)\,
T_{1}(u+2\eta+i\pi),\label{Fused-Mon-matrix-3}\\
&&P^{(3)}_{12} \hat{T}_1(u)\,\hat{T}_2(u+4\eta)P^{(1)}_{12}=\prod_{l=1}^N \rho_3(u+\theta_l)\,
\hat{T}_{1}(u+2\eta+i\pi),\label{Fused-Mon-matrix-4}
\eea where $\{\theta_l|l=1,\ldots,N\}$ are the inhomogeneous parameters defined in (\ref{Mon-V-1})-(\ref{Mon-V-2}).

Now let us construct the corresponding fusion procedures for the $K$-matrices following \cite{Mez92,Zho96}. The properties (\ref{Fusion}) of the
$R$-matrix at the degenerate points  and the RE (\ref{RE-V}) and its dual RE (\ref{DRE-V}) allow
us to construct the fused $K$-matrices from the origin ones (\ref{K-matrix-1}) and (\ref{K-matrix-2}). After some tedious calculations,
we find that
\bea
&&\hspace{-1.2truecm}P^{(1)}_{21}K^-_1(u)R_{21}(2u+6\eta+i\pi)K_2^-(u+6\eta+i\pi)P^{(1)}_{12}={\rm Det}_q(K^-(u))P^{(1)}_{21},\label{Fused-K-1}\\[6pt]
&&\hspace{-1.2truecm}P^{(1)}_{12}K^+_2(u\hspace{-0.04truecm}+\hspace{-0.04truecm}6\eta\hspace{-0.04truecm}+\hspace{-0.04truecm}i\pi)
   \hspace{-0.08truecm}{\cal M}_1\hspace{-0.08truecm}
   R_{12}\hspace{-0.02truecm}(\hspace{-0.04truecm}-\hspace{-0.04truecm}2u\hspace{-0.04truecm}+\hspace{-0.04truecm}6\eta
   \hspace{-0.04truecm}+\hspace{-0.04truecm}i\pi)
   \hspace{-0.08truecm}{\cal M}_1^{-1}\hspace{-0.08truecm}K_1^+(u)\hspace{-0.08truecm}P^{(1)}_{21}\hspace{-0.02truecm}=
   \hspace{-0.02truecm}{\rm Det}_q(K^+(u))P^{(1)}_{12},\label{Fused-K-2}\\[6pt]
&&\hspace{-1.2truecm}P^{(3)}_{21}K^-_1(u)R_{21}(2u+4\eta)K_2^-(u+4\eta)P^{(3)}_{12}=f_-(u)\,K_1^-(u+2\eta+i\pi),\label{Fused-K-3}\\[6pt]
&&\hspace{-1.2truecm}P^{(3)}_{12}K^+_2(u+4\eta){\cal M}_1R_{12}(-2u+8\eta){\cal M}_1^{-1}K_1^+(u)P^{(3)}_{21}=f_+(u)\,K^+_1(u+2\eta+i\pi),\label{Fused-K-4}
\eea where the functions $f_{\pm}(u)$ and ${\rm Det}_q(K^{\pm}(u))$ are
\bea
\hspace{-1.2truecm}f_-(u)&=&-2(1-2e^{-\e}\sinh(u-\eta))\cosh(u-\eta)\sinh(u+4\eta),\label{F-function-1}\\
\hspace{-1.2truecm}f_+(u)&=&2(1-2e^{-\e'}\sinh(u-\eta))\cosh(u-\eta)\sinh(u-6\eta),\label{F-function-2}\\[6pt]
\hspace{-1.2truecm}{\rm Det}_q(K^-(u))&=&tr_{12}\lt\{P^{(1)}_{21}K^-_1(u)R_{21}(2u+6\eta+i\pi)K_2^-(u+6\eta+i\pi)P^{(1)}_{12}\rt\}\no\\[6pt]
        &=&-2(1-2e^{-\e}\sinh(u-\eta))(1+e^{-\e}\sinh(u+\eta))\no\\[6pt]
        &&\quad\quad\times\sinh(u+6\eta)\cosh(u+\eta),
        \label{Det-1}\\[6pt]
\hspace{-1.2truecm}{\rm Det}_q(K^+(u))&=&tr_{12}\hspace{-0.08truecm}\lt\{\hspace{-0.08truecm}
         P^{(1)}_{12}\hspace{-0.02truecm}K^+_2\hspace{-0.02truecm}(u\hspace{-0.02truecm}+\hspace{-0.02truecm}6\eta
         \hspace{-0.02truecm}+\hspace{-0.02truecm}i\pi){\cal M}_1\hspace{-0.08truecm}R_{12}\hspace{-0.02truecm}(\hspace{-0.02truecm}-\hspace{-0.02truecm}2u
         \hspace{-0.02truecm}+\hspace{-0.02truecm}6\eta\hspace{-0.02truecm}+\hspace{-0.02truecm}i\pi)
         \hspace{-0.02truecm}{\cal M}_1^{-1}\hspace{-0.08truecm}K_1^+(u)\hspace{-0.08truecm}P^{(1)}_{21}\hspace{-0.08truecm}\rt\}\no\\[6pt]
        &=&2(1-2e^{-\e'}\sinh(u-\eta))(1+e^{-\e'}\sinh(u+\eta))\no\\[6pt]
        &&\quad\quad\times\sinh(u-6\eta)\cosh(u-\eta).
        \label{Det-2}
\eea Following the method in \cite{Cao13} and using the relations (\ref{Fused-Mon-matrix-1})-(\ref{Fused-K-4}), after a long calculation, we
find the transfer matrix given by (\ref{trans}) satisfies the following operator identities
\bea
t(\pm\theta_j)t(\pm\theta_j+6\eta+i\pi)&=&\frac{\delta_1(u)\,\times{\rm id}}
            {\rho_1(2u)}\lt|_{u=\pm\theta_j}\rt.,\quad  j=1,\ldots,N,\label{Id-1}\\[6pt]
t(\pm\theta_j)t(\pm\theta_j+4\eta)&=&\frac{\delta_2(u)\,\times t(u+2\eta+i\pi)}
            {\rho_2(-2u+8\eta)}\lt|_{u=\pm\theta_j}\rt.,\quad j=1,\ldots,N.\label{Id-2}
\eea Here the functions $\delta_1(u)$ and $\delta_2(u)$ are
\bea
\delta_1(u)&=&{\rm Det}_q(K^-(u)) {\rm Det}_q(K^+(u))\prod_{l=1}^N\rho_1(u-\theta_l)\rho_1(u+\theta_l),\label{Delta-1}\\
\delta_2(u)&=&f_-(u)f_+(u)\prod_{l=1}^N\rho_3(u-\theta_l)\rho_3(u+\theta_l).\label{Delta-2}
\eea
Following the method in \cite{Yan08} and using the the explicit expressions (\ref{K-matrix-1}) and (\ref{K-matrix-2}) of the $K$-matrices, we
find that the transfer matrix (\ref{trans}) possesses the crossing symmetry
\bea
t(u)=t(-u+6\eta+i\pi).\label{trans-crossing}
\eea The proof of the above relation is given in Appendix A. We remark that the very operator identities (\ref{Id-1})-(\ref{Id-2}) at the points $-\theta_j$
is guaranteed by their equalities at the points $\theta_j$ and the crossing symmetry (\ref{trans-crossing}) of the transfer matrix. Therefore, in the following
we will use the operator identities  (\ref{Id-1})-(\ref{Id-2}) at the points $\theta_j$ and the crossing symmetry (\ref{trans-crossing}) to determine the
eigenvalues of the transfer matrix.

Now let us derive some properties of the transfer matrix. Firstly the expressions of the $K$-matrices $K^{\pm}(u)$ given by (\ref{K-matrix-1}) and
(\ref{K-matrix-2}) imply that $K^{\pm}(u)$ have the following periodicity
\bea
K^{\pm}(u+2i\pi)=K^{\pm}(u).\no
\eea This leads to the periodicity of the transfer matrix $t(u)$
\bea
t(u+2i\pi)=t(u).\label{Perodicity-trans}
\eea Moreover the explicit expressions of the $R$-matrix  and the $K$-matrices allow us to have the following asymptotic
behaviors of the transfer matrix
\bea
\lim_{u\rightarrow \pm\infty}t(u)=(\frac{1}{2})^{2N}e^{-\e-\e'}e^{\pm 2(N+1)(u-3\eta)}(1+2\cosh(\sigma'-\sigma+2\eta))\,\times {\rm id}+\ldots.
\label{Asymptotic}
\eea  The explicit expressions of the $K$-matrices given by (\ref{K-matrix-1}) and
(\ref{K-matrix-2}) also imply that $K^{\pm}(u)$ satisfy the following properties
\bea
&&K^-(0)=(1+2e^{-\e}\sinh\eta),\quad K^+(6\eta+i\pi)=(1+2e^{-\e'}\sinh\eta){\cal M},\\[6pt]
&&K^-(i\pi)=(1-2e^{-\e}\sinh\eta),\quad K^+(6\eta)=(1-2e^{-\e'}\sinh\eta){\cal M}.
\eea The unitarity (\ref{Unitarity}) and the crossing unitarity (\ref{crossing-unitarity}) of the $R$-matrix imply that the one-row
monodromy matrices $T(u)$ given by (\ref{Mon-V-1}) and $\hat{T}(u)$  given by (\ref{Mon-V-2}) satisfy the following relations
\bea
&&T_0(u)\hat{T}_0(-u)=\prod_{l=1}^N\rho_1(u-\theta_l)\times {\rm id},\\
&&T^{t_0}_0(u){\cal M}_0\hat{T}^{t_0}_0(-u+12\eta){\cal M}_0^{-1}=\prod_{l=1}^N\rho_2(u-\theta_l)\times {\rm id}.\label{T-Relation}
\eea The above relations allow us to work out the transfer matrix at special points $0,\,i\pi,\,6\eta,\,6\eta+i\pi$ as follows
\bea
&&t(0)=t(6\eta+i\pi)=(1+2e^{-\e}\sinh\eta)tr\{K^+(0)\}\prod_{l=1}^N\rho_1(-\theta_l)\times {\rm id},\label{t-value-1}\\
&&t(i\pi)=t(6\eta)=(1-2e^{-\e}\sinh\eta)tr\{K^+(i\pi)\}\prod_{l=1}^N\rho_1(i\pi-\theta_l)\times {\rm id}.\label{t-value-2}
\eea


\section{ Functional relations and the T-Q relation}
\label{T-Q} \setcounter{equation}{0}

The property $[t(u), t(v)]=0$ implies that the eigenstates of $t(u)$ are independent of $u$. Suppose $|\Psi\rangle$ is an eigenstate of
$t(u)$ with an eigenvalue $\Lambda(u)$, namely,
\bea
t(u)|\Psi\rangle =\Lambda(u)|\Psi\rangle.\no
\eea
The very operator identities (\ref{Id-1})-(\ref{Id-2}) of the transfer matrix at the points $\theta_j$  imply the corresponding eigenvalue
$\Lambda(u)$  satisfies the similar relations
\bea
&&\Lambda(\theta_j)\Lambda(\theta_j+6\eta+i\pi)=\frac{\delta_1(u)}
            {\rho_1(2u)}\lt|_{u=\theta_j}\rt.,\quad
            j=1,\ldots,N,\label{Eigen-Id-1}\\[6pt]
&&\Lambda(\theta_j)\Lambda(\theta_j+4\eta)=\frac{\delta_2(u)\,\times \Lambda(u+2\eta+i\pi)}
            {\rho_2(-2u+8\eta)}\lt|_{u=\theta_j}\rt.,\quad j=1,\ldots,N.\label{Eigen-Id-2}
\eea
where the functions $\delta_i(u)$ are given by (\ref{Delta-1}) and (\ref{Delta-2}). The crossing symmetry (\ref{trans-crossing}) of the transfer
matrix leads to the crossing symmetry of its eigenvalues
\bea
\L(u)=\L(-u+6\eta+i\pi).\label{Eigen-crossing}
\eea
The properties
of the transfer matrix $t(u)$ given by (\ref{Perodicity-trans}), (\ref{Asymptotic}) and
(\ref{t-value-1})-(\ref{t-value-2}) imply that the corresponding eigenvalue
$\Lambda(u)$ satisfies the following relations:
\bea
&&\L(u+2i\pi)=\L(u),\label{Per-Eign}\\[6pt]
&&\L(0)=(1+2e^{-\e}\sinh\eta)tr\{K^+(0)\}\prod_{l=1}^N\rho_1(-\theta_l),\label{Eigen-value-1}\\[6pt]
&&\L(i\pi)=(1-2e^{-\e}\sinh\eta)tr\{K^+(i\pi)\}\prod_{l=1}^N\rho_1(i\pi-\theta_l),\label{Eigen-value-2}\\[6pt]
&&\lim_{u\rightarrow \pm\infty}\L(u)=(\frac{1}{2})^{2N}e^{-\e-\e'}e^{\pm 2(N+1)(u-3\eta)}(1+2\cosh(\sigma'-\sigma+2\eta))+\ldots.
\label{Eigen-Asymptotic}
\eea The periodicity (\ref{Per-Eign}) and the asymptotic behavior (\ref{Eigen-Asymptotic}) of the eigenvalue $\L(u)$,
the analyticity of the R-matrix and K-matrices and the $u$-free eigenstate
lead to the fact  that the eigenvalue $\Lambda(u)$ further possesses
the property
\bea \L(u) \mbox{, as an entire function of $u$, is a
trigonometric polynomial of degree $2N+2$}.\label{Eigen-Anal}
\eea
Namely, $\L(u)$ is a Laurent polynomial of
$e^u$ with $4(N+1)+1$ unknown coefficients. The crossing relation (\ref{Eigen-crossing}) reduces the number of the independent unknown
coefficients to $2(N+1)+1$. Thus one needs $2(N+1)+1$ conditions to determine these coefficients. Therefore the  relations  (\ref{Eigen-Id-1})-(\ref{Eigen-Anal})  completely characterize  the spectrum of the transfer matrix of the model.

\subsection{T-Q ansatz}
Based on the properties (\ref{Eigen-Id-1})-(\ref{Eigen-Anal}) of the eigenvalues $\L(u)$, following the ODBA method
developed in \cite{Cao1,Cao13}, let us propose  the following conjecture for the eigenvalues,
\bea
\L(u)&=&\prod_{l=1}^Nc(u-\theta_l)c(u+\theta_l)(1-2e^{-\e}\sinh(u-\eta))(1-2e^{-\e'}\sinh(u-\eta))\no\\[6pt]
&&\quad\quad \times \frac{\sinh(u-6\eta)\cosh(u-\eta)}{\sinh(u-2\eta)\cosh(u-3\eta)}\frac{Q_1(u+4\eta)}{Q_2(u)}\no\no\\[6pt]
&&+\prod_{l=1}^Nd(u-\theta_l)d(u+\theta_l)(1-2e^{-\e}\sinh(u-5\eta))(1-2e^{-\e'}\sinh(u-5\eta))\no\\[6pt]
&&\quad\quad\times\frac{\sinh u\cosh(u-5\eta)}{\sinh(u-4\eta)\cosh(u-3\eta)}\frac{Q_2(u-6\eta+i\pi)}{Q_1(u-2\eta+i\pi)}\no\\[6pt]
&&+\prod_{l=1}^Nb(u-\theta_l)b(u+\theta_l)(1+2e^{-\e}\sinh(u-3\eta))(1+2e^{-\e'}\sinh(u-3\eta))\no\\[6pt]
&&\quad\quad  \times\frac{\sinh
u\sinh(u-6\eta)}{\sinh(u-2\eta)\sinh(u-4\eta)}
\frac{Q_1(u+2\eta+i\pi)Q_2(u-4\eta)}{Q_2(u-2\eta+i\pi)Q_1(u)}\no\\[6pt]
&&+ 4^{1-N}c\, \frac{\sinh
u\sinh(u-6\eta)}{\cosh(u-3\eta)}\prod_{l=1}^Nc(u-\theta_l)c(u+\theta_l)d(u-\theta_l)d(u+\theta_l)\no\\[6pt]
&&\qquad\times\left[\frac{Q_1(u+2\eta+i\pi)\sinh^{l_1}(u-3\eta)\sinh^{l_1}(u-\eta)\cosh^{l_2}(u-2\eta)}{Q_1(u)Q_2(u)}
\right.\no \\[6pt]
&&\qquad\qquad \left. -(-1)^{l_2}
\frac{Q_2(u\hspace{-0.08truecm}-\hspace{-0.08truecm}4\eta)
\sinh^{l_1}(u\hspace{-0.08truecm}-\hspace{-0.08truecm}3\eta)
\sinh^{l_1}(u\hspace{-0.08truecm}-\hspace{-0.08truecm}5\eta)
\cosh^{l_2}(u\hspace{-0.08truecm}-\hspace{-0.08truecm}4\eta)}{Q_1(u-2\eta+i\pi)Q_2(u-2\eta+i\pi)}
\right], \label{T-Q-Main}\eea
where the functions $c(u)$, $b(u)$ and $d(u)$ are the elements of the $R$-matrix given by (\ref{R-element-2})
and the functions  $Q_i(u)$ and the constant $c$ are given by
\bea && Q_1(u)=\prod_{k=1}^{\bar N} \sinh[\frac{u-\l_k}{2}-\eta],\quad \bar N=4(N+l_1)+2(l_2-1),\label{Q1} \\[6pt]
&& Q_2(u)=\prod_{k=1}^{\bar N}\sinh[\frac{u+\l_k}{2}-\eta],\label{Q2}\\[6pt]
&&c=(-1)^{(l_2-1)}e^{-\e-\e'}\lt\{\frac{\cosh(\sigma'-\sigma+2\eta)-\cosh(\d_{l_1,l_2}\eta-\sum_{j=1}^{\bar N}\l_j)}
{\cosh(\frac{\d_{l_1,l_2}\eta}{2}-\frac{1}{2}\sum_{j=1}^{\bar N}\l_j)}\rt\},\label{c-const}\\[6pt]
&&\delta_{l_1,l_2}=4N+8l_1+4l_2-2.\label{Integer}
 \eea Here $l_1$ and $l_2$ are arbitrary non-negative integers \footnote{The results of anisotropic
spin-$\frac12$ chains \cite{Cao1} strongly suggest that fixed $l_1$ and $l_2$ might give a complete set of eigenvalues of the transfer matrix. In such a sense, different $l_1$ and $l_2$ might only give different parameterizations of the eigenvalues but not different states.}.
It is easy to check that the functions $Q_i(u)$ possess the following properties
\bea
Q_i(u+2i\pi)=Q_i(u),\quad {\rm for}\,\, i=1,2,\quad {\rm and}\quad Q_2(u)=Q_1(-u+4\eta).\label{Q-property}
\eea With the help of the above relations, we have checked that the $T-Q$ ansatz (\ref{T-Q-Main}) indeed satisfies the relations (\ref{Eigen-crossing})-(\ref{Eigen-value-2}).
Moreover the special choice of the constant $c$ given by (\ref{c-const}) implies that the ansatz also satisfies the asymptotic behavior (\ref{Eigen-Asymptotic}).
The explicit expressions of the functions $b(u)$, $c(u)$ and $d(u)$ given by (\ref{R-matrix})-(\ref{R-element-2})
imply that
\bea
b(0)=d(0)=c(6\eta+i\pi)=b(6\eta+i\pi)=c(4\eta)=d(2\eta+i\pi)=0.
\eea  These properties give rise to the fact that the $T-Q$ ansatz (\ref{T-Q-Main})  satisfies the very functional relations (\ref{Eigen-Id-1})-(\ref{Eigen-Id-2}).
This means that our ansatz (\ref{T-Q-Main}) actually satisfies the relations (\ref{Eigen-Id-1})-(\ref{Eigen-Asymptotic}) except the analytic properties (\ref{Eigen-Anal}).

From the explicit expression (\ref{T-Q-Main}) of $\L(u)$, one may find that the $T-Q$ ansatz (\ref{T-Q-Main}) might have some apparent simple
poles at the following points:
\bea
2\eta,\,\,4\eta,\,\,3\eta+i\frac{\pi}{2},\quad {\rm mod}(i\pi),\label{pole-1}
\eea  and
\bea
\l_j+2\eta,\,\,-\l_j+2\eta,\,\, \l_j+4\eta+i\pi,\,\,-\l_j+4\eta+i\pi,\quad {\rm mod} (2i\pi),\quad j=1,\ldots,\bar N.\label{pole-2}
\eea Direct calculation shows that the residues of the $T-Q$ ansatz (\ref{T-Q-Main}) at the points given by  (\ref{pole-1}) vanishes,
or the ansatz has no singularity at these points. Moreover we have checked that  the $T-Q$ ansatz (\ref{T-Q-Main}) also has no singularity
at the points given by  (\ref{pole-2}) provided that the $\bar{N}$ parameters $\{\l_j|j=1,\ldots,\bar{N}\}$ satisfy the following
BAEs
\bea
&&\hspace{-1.2truecm}\frac{(1+2e^{-\e}\sinh(\l_j-\eta))(1+2e^{-\e'}\sinh(\l_j-\eta))\cosh(\l_j-\eta)Q_2(\l_j-2\eta)Q_2(\l_j+2\eta)}
{4\sinh\l_j\sinh(\l_j-2\eta)\sinh^{l_1}(\l_j+\eta)\sinh^{l_1}(\l_j-\eta)\cosh^{l_2}\l_j}
\no\\
&&\hspace{-1.2truecm}\quad =
\hspace{-0.04truecm}-\hspace{-0.04truecm}c\,Q_2(\l_j\hspace{-0.04truecm}+\hspace{-0.04truecm}i\pi)
\prod_{l=1}^N\sinh[\frac{\l_j\hspace{-0.04truecm}-\hspace{-0.04truecm}\theta_l}{2}\hspace{-0.04truecm}-\hspace{-0.04truecm}\eta]
\sinh[\frac{\l_j\hspace{-0.04truecm}+\hspace{-0.04truecm}\theta_l}{2}\hspace{-0.04truecm}-\hspace{-0.04truecm}\eta]
\cosh[\frac{\l_j\hspace{-0.04truecm}-\hspace{-0.04truecm}\theta_l}{2}]
\cosh[\frac{\l_j\hspace{-0.04truecm}+\hspace{-0.04truecm}\theta_l}{2}],\no\\[6pt]
&&\quad\quad\quad\quad j=1,\ldots,\bar{N},\label{BAE-1}
\eea where the function $Q_2(u)$ is given by (\ref{Q2}). The non-singular property of
the $T-Q$ ansatz (\ref{T-Q-Main}) at the points (\ref{pole-2}) can be verified by directly calculating the residues of
the ansatz at these points.

Finally we conclude that the $T-Q$ ansatz (\ref{T-Q-Main}) indeed satisfies (\ref{Eigen-Id-1})-(\ref{Eigen-Anal}) as it is required, if the
the $\bar{N}$ parameters $\{\l_j|j=1,\ldots,\bar{N}\}$ satisfy the associated BAEs (\ref{BAE-1}). Thus the $\L(u)$ given by (\ref{T-Q-Main}), when
the $\bar{N}$ parameters $\{\l_j|j=1,\ldots,\bar{N}\}$ satisfy the associated BAEs (\ref{BAE-1}), becomes the eigenvalue
of the transfer matrix $t(u)$ given by (\ref{trans}) for the Izergin-Korepin model with the general non-diagonal boundary terms specified by
the non-diagonal $K$-matrices $K^{\pm}(u)$ given by (\ref{K-matrix-1}) and (\ref{K-matrix-2}).

\subsection{Reduction to the conventional $T-Q$ ansatz}

It follows from (\ref{c-const}) that the parameter $c$ does depend on not
only the boundary parameters but also the parameters $\{\l_j\}$ (such
a dependence    also appeared in the anisotropic spin-$\frac{1}{2}$ chains with arbitrary boundary fields \cite{Cao1}). The vanishing
condition of $c$, i.e. $c=0$, will lead to the constraint (see (\ref{constraint-0}) below) among the boundary parameters, in this case one might find a proper
``local vacuum" to apply the conventional Bethe ansatz \cite{Cao03,Yan07}. The Bethe ansatz equations
(\ref{BAE-1}) imply that for this case the parameters $\{\l_j\}$ have to form
two types of pairs:
\bea (\l_j,-\l_j),\quad\quad
(\l_j,-\l_j+4\eta).\label{Pairs}
\eea Suppose  the number of the first type
pairs is $M$ ($M$ being a non-negative integer such that $0\leq M\leq \frac{\bar{N}}{2}$), the resulting $T-Q$ relation (\ref{T-Q-Main}) becomes the
conventional one \cite{Bax82}
\bea
\L(u)&=&\prod_{l=1}^Nc(u-\theta_l)c(u+\theta_l)(1-2e^{-\e}\sinh(u-\eta))(1-2e^{-\e'}\sinh(u-\eta))\no\\
&&\quad\quad \times \frac{\sinh(u-6\eta)\cosh(u-\eta)}{\sinh(u-2\eta)\cosh(u-3\eta)}\frac{Q(u+4\eta)}{Q(u)}\no\\
&&+\prod_{l=1}^Nd(u-\theta_l)d(u+\theta_l)(1-2e^{-\e}\sinh(u-5\eta))(1-2e^{-\e'}\sinh(u-5\eta))\no\\
&&\quad\quad\times\frac{\sinh u\cosh(u-5\eta)}{\sinh(u-4\eta)\cosh(u-3\eta)}\frac{Q(u-6\eta-i\pi)}{Q(u-2\eta+i\pi)}\no\\
&&+\prod_{l=1}^Nb(u-\theta_l)b(u+\theta_l)(1+2e^{-\e}\sinh(u-3\eta))(1+2e^{-\e'}\sinh(u-3\eta))\no\\
&&\quad\quad  \times\frac{\sinh
u\sinh(u-6\eta)}{\sinh(u-2\eta)\sinh(u-4\eta)}
\frac{Q(u+2\eta+i\pi)Q(u-4\eta)}{Q(u-2\eta+i\pi)Q(u)},\label{T-Q-1}
\eea where the resulting function $Q(u)$ is
\bea
Q(u)=\prod_{j=1}^M\sinh[\frac{u-\l_j}{2}-\eta] \sinh[\frac{u+\l_j}{2}-\eta].\label{Q3}
\eea The resulting BAEs become
\bea
&&\hspace{-1.2truecm}\prod_{l=1}^N\frac{\sinh[\frac{\l_j-\theta_l}{2}-\eta]\sinh[\frac{\l_j+\theta_l}{2}-\eta]}
{\sinh[\frac{\l_j-\theta_l}{2}+\eta]\sinh[\frac{\l_j+\theta_l}{2}+\eta]}
\frac{(1-2e^{-\e}\sinh(\l_j+\eta))(1-2e^{-\e'}\sinh(\l_j+\eta))}
{(1+2e^{-\e}\sinh(\l_j-\eta))(1+2e^{-\e'}\sinh(\l_j-\eta))}\no\\[6pt]
&&\hspace{-1.2truecm}\quad =-\frac{\sinh(\l_j+2\eta)\cosh(\l_j-\eta)}{\sinh(\l_j-2\eta)\cosh(\l_j+\eta)}
\frac{Q(-\l_j-2\eta)Q(-\l_j+4\eta+i\pi)}{Q(\l_j-2\eta)Q(\l_j+4\eta+i\pi)},\,\,j=1,\ldots,M.\label{BAE-2}
\eea
On the other hand, the form of the pairs (\ref{Pairs}) implies that
\bea
\sum_{j+1}^{\bar N}\l_j=\sum_{j=1}^{\frac{\bar N-2M}{2}}(\l_j-\l_j+4\eta)=(2\bar N-4M)\eta. \no
\eea The relation (\ref{c-const}) and the definition (\ref{Q1}) of $\bar N$   give rise to  the following
constraint between the boundary parameters \footnote{One can always choose the non-negative integers $l_1$ and $l_2$
such that $M$ becomes any non-negative integer since that $0\leq M\leq 2(N+l_1)+l_2-1=\bar N/2$.}
\bea
\cosh(\sigma'-\sigma+2\eta)=\cosh(4M\eta-4N\eta+2\eta),\quad 0\leq M.\label{constraint-0}
\eea

\subsubsection{For a generic $\eta$}
For a generic $\eta$, there exist three solutions to the constraint (\ref{constraint-0}). If
the boundary parameters $\sigma$ and $\sigma'$ satisfy the constraint
\bea
 \sigma'-\sigma=-4k\eta\,\, {\rm mod}(2i\pi), \,\,k\leq -N, \quad {\rm and}\,\,k\in Z,\label{constraint-1}
\eea the corresponding $M$ can take only one allowed value,
\bea
M=N-k.\label{M-values-1}
\eea If the boundary parameters $\sigma$ and $\sigma'$ satisfy the following constraint
\bea
\sigma'-\sigma=-4k\eta\,\, {\rm mod}(2i\pi),\quad 1-N\leq k\leq N, \quad {\rm and}\,\,k\in Z,\label{constraint-2}
\eea then $M$ can  take the two allowed values denoted by $M^{\pm}$ respectively,
\bea
M^-=N-k,\quad M^+=N+k-1.\label{M-value-2}
\eea
If the boundary parameters $\sigma$ and $\sigma'$ satisfy
\bea
 \sigma'-\sigma=-4k\eta\,\, {\rm mod}(2i\pi),\,\, N+1\leq k, \quad {\rm and}\,\,k\in Z,\label{constraint-3}
\eea  $M$ can take only one allowed value,
\bea
M=N+k-1.\label{M-values-3}
\eea For the case that  the boundary parameters satisfy the relation (\ref{constraint-1}) (or (\ref{constraint-3})), the eigenvalue
of the transfer matrix is characterized by a fixed $M$ in (\ref{M-values-1})
(or (\ref{M-values-3})). This $\L(u)$ itself might give the complete set of the eigenvalues of the transfer matrix. However, if the boundary
parameters obey the relation (\ref{constraint-2}), there exist two $\L^{\pm}(u)$ corresponding to the two different
allowed $M$ in (\ref{M-value-2}). Similar as that of the anisotropic spin-$\frac{1}{2}$ chain with arbitrary boundary fields
\cite{Cao03,Nep03-1,Nep03-2,Yan06}, these two $\L^{\pm}(u)$ together might constitute the complete set of the eigenvalues of the transfer matrix.

\subsubsection{For some degenerate  $\eta$}

Similar as that in the closed XYZ chain and the anisotropic spin-$\frac{1}{2}$ chain with
arbitrary boundary fields \cite{Cao1}, if the isotropic (or crossing) parameter $\eta$ takes
the following discrete value
\bea
\eta=\frac{\s-\s'}{4N-4M}+\frac{2i\pi m}{4N-4M},\quad m,M\in Z,\,\, {\rm and} \,\, 0\leq M,\label{eta-value}
\eea   the constraint (\ref{constraint-0}) is automatically satisfied for arbitrary boundary parameters $\e$,
$\e'$, $\sigma$ and $\sigma'$. Thus in this case the eigenvalue of the transfer matrix is given by
(\ref{T-Q-1}) and the corresponding BAEs are given by (\ref{BAE-2}). It should be emphasized that these degenerate points
(\ref{eta-value}) become
dense in the  thermodynamic limit
($N,m,M\rightarrow \infty$). This enables one to obtain the
thermodynamic properties (up to the order of $O(N^{-2})$) of the
model for generic values of $\eta$ via the conventional
thermodynamic Bethe ansatz methods \cite{yb,Tak99}. This method has been
proven to be very successful in the derivation of the surface energy of the
XXZ spin chain with arbitrary boundary fields \cite{Li14-1}.


\section{Reduction to the case with diagonal $K$-matrices }
\label{Diag} \setcounter{equation}{0}

Now let us  consider the diagonal $K$-matrices (i.e., taking the limits $\e,\e'\rightarrow +\infty$ of (\ref{K-matrix-1}) and (\ref{K-matrix-2})).
The resulting $K$-matrices read
\bea
K^-(u)={\rm id},\quad K^+(u)=\cal M,\label{Diagonal-Kmatrix}
\eea where the matrix $\cal M$ is given by (\ref{M-matrix}).
The $T-Q$ ansatz (\ref{T-Q-1}) is reduced to the one \cite{Mez92} obtained by analytic Bethe ansatz method
\bea
\L(u)&=&\prod_{l=1}^Nc(u-\theta_l)c(u+\theta_l)\,\frac{\sinh(u-6\eta)\cosh(u-\eta)}{\sinh(u-2\eta)\cosh(u-3\eta)}\frac{Q(u+4\eta)}{Q(u)} \no\\
&&+\prod_{l=1}^Nd(u-\theta_l)d(u+\theta_l)\,\frac{\sinh u\cosh(u-5\eta)}{\sinh(u-4\eta)\cosh(u-3\eta)}\frac{Q(u-6\eta+i\pi)}{Q(u-2\eta+i\pi)} \no\\
&&+\prod_{l=1}^Nb(u-\theta_l)b(u+\theta_l)\,\frac{\sinh u\sinh(u-6\eta)}{\sinh(u-2\eta)\sinh(u-4\eta)}\no\\
&&\quad\quad\times\frac{Q(u-4\eta)Q(u+2\eta+i\pi)}{Q(u-2\eta+i\pi)Q(u)},
\label{T-Q-2}
\eea where the function $Q(u)$ is still given by (\ref{Q3}). In particular, in this case the vanishing condition of $c$  is automatically
satisfied thanks to the relation (\ref{c-const}). In this case the $U(1)$-symmetry is recovered and $M$ can take any of the following
values \footnote{The allowed values (\ref{M-value-4}) of $M$  can be deduced by  studying  the
asymptotic behavior of the eigenvalues of transfer matrix \cite{Mez92}.},
\bea
M=0,1,\ldots, 2N.\label{M-value-4}
\eea For each of the above allowed values, the resulting BAEs become
\bea
&&\prod_{l=1}^N\frac{\sinh[\frac{\l_j-\theta_l}{2}-\eta]\sinh[\frac{\l_j+\theta_l}{2}-\eta]}
{\sinh[\frac{\l_j-\theta_l}{2}+\eta]\sinh[\frac{\l_j+\theta_l}{2}+\eta]}
\frac{\sinh(\l_j-2\eta)\cosh(\l_j+\eta)}{\sinh(\l_j+2\eta)\cosh(\l_j-\eta)}\no\\[6pt]
&&\quad\quad =-\frac{Q(-\l_j-2\eta)Q(-\l_j+4\eta+i\pi)}{Q(\l_j-2\eta)Q(\l_j+4\eta+i\pi)},
\,\,j=1,\ldots,M.\label{BAE-3}
\eea


\section{Conclusions}
\label{Con} \setcounter{equation}{0}

The Izergin-Korepin model with general non-diagonal boundary terms specified by the most general
non-diagonal $K$ matrices given by (\ref{K-matrix-1}) and (\ref{K-matrix-2}) has been studied by
the off-diagonal Bethe ansatz method. Based on some intrinsic properties of the $R$-matrix and $K$-matrices, we derive the very
functional relations (\ref{Id-1}) and (\ref{Id-2}) of the transfer matrix. These relations, together with other properties,
allow us to construct an off-diagonal (or inhomogeneous) $T-Q$ relation (\ref{T-Q-Main}) of the eigenvalue of
the transfer matrix and the associated BAEs (\ref{BAE-1}). When the boundary parameters satisfy one
constraint (\ref{constraint-0}), the resulting $T-Q$ relation is reduced to the conventional one
(\ref{T-Q-1}), which might allow one to use the method developed in \cite{Li14-1} to study  the
thermodynamic properties (up to the order of $O(N^{-2})$) of the
model for generic values of $\eta$ via the conventional
thermodynamic Bethe ansatz methods \cite{yb,Tak99}. Taking the
limit $\e,\e'\rightarrow +\infty$, the corresponding $K$-matrices become diagonal ones and the
resulting $T-Q$ relation is then reduced  to that in \cite{Mez92}.

\section*{Acknowledgments}

The financial support from  the National Natural Science Foundation
of China (Grant Nos. 11174335, 11031005, 11375141, 11374334), the
National Program for Basic Research of MOST (973 project under grant
No.2011CB921700), the State Education Ministry of China (Grant No.
20116101110017) and BCMIIS are gratefully acknowledged.


\section*{Appendix A: Proof of the crossing symmetry }
\setcounter{equation}{0}
\renewcommand{\theequation}{A.\arabic{equation}}
Direct
calculation shows that the $K$-matrices given by (\ref{K-matrix-1})-(\ref{K-matrix-2})
satisfy the following relations
\bea
&&K^-(u)K^-(-u)=\Delta_{-}(u)\times {\rm id},\label{K-unitarity-1}\\[6pt]
&&V^{t}\,K^+(-u+6\eta+i\pi)\,VV^t\,K^+(u+6\eta+i\pi)V= \Delta_{+}(u)\times {\rm id},\label{K-unitarity-2}
\eea where the matrix $V$ is given by (\ref{V-matrix}) and the functions $\Delta_{\pm}(u)$ are
\bea
\Delta_{-}(u)&=&(1-2e^{-\e}\sinh(u-\eta))(1+2e^{-\e}\sinh(u+\eta)),\label{D-1}\\[6pt]
\Delta_{+}(u)&=&(1-2e^{-\e'}\sinh(u-\eta))(1+2e^{-\e'}\sinh(u+\eta)).\label{D-2}
\eea Similar as those in \cite{Yan08}, let us introduce the following matrices $\bar{K}^{\pm}(u)$
\bea
 &&\bar{K}_1^-(u)=tr_2\lt\{P_{12}\,R_{21}(-2u)\,V_2\,{K_2^-}^{t_2}(u+6\eta+i\pi)\,V_2^{t_2}\rt\},\\[6pt]
 &&\bar{K}_1^+(u+6\eta+i\pi)=tr_2\lt\{P_{12}\,R_{12}(2u)\,{K_2^+}^{t_2}(u)\rt\}.
\eea Substituting the expressions (\ref{K-matrix-1})-(\ref{K-matrix-2}) of the $K$-matrices into the above relations, we have that
the $K$-matrices satisfy the following crossing relation
\bea
&&\bar{K}^-(u)=\frac{{\rm Det}_q(K^-(u))}{\Delta_-(u)}\,K^-(-u),\label{K-crossing-1}\\[6pt]
&&\bar{K}^+(u+6\eta+i\pi)=\frac{{\rm Det}_q(K^+(u))}{\Delta_+(u)}\,V^t\,K^+(-u+6\eta+i\pi)\,V,\label{K-crossing-2}
\eea where the functions ${\rm Det}(K^-(u))$ and ${\rm Det}(K^+(u))$ are given respectively by (\ref{Det-1}) and (\ref{Det-2}).
Combining the crossing relation (\ref{crosing}) and PT-symmetry (\ref{PT}) of the $R$-matrix, one may derive the following relations
\bea
&&R_{21}(u)=V_1^{t_1}\,R_{12}^{t_1}(-u+6\eta+i\pi)\,V_1^{t_1}, \\[6pt]
&&R_{21}(u)=V_2\,R_{12}^{t_2}(-u+6\eta+i\pi)\,V_2.
\eea  Then the above relations, (\ref{crosing}) and the expression (\ref{V-matrix}) give rise to the following relation between the one-row
monodromy matrices $T(u)$ and $\hat{T}(u)$ give by (\ref{Mon-V-1}) and (\ref{Mon-V-2})
\bea
&&T_0^{t_0}(-u+6\eta+i\pi)=V_0^{t_0}\,\hat{T}_0(u)\,V_0^{t_0},\\[6pt]
&&\hat{T}_0^{t_0}(-u+6\eta+i\pi)=V_0\,T_0(u)\,V_0.
\eea Following the method in \cite{Yan08} and using the QYBE (\ref{QYB}), we can prove that the transfer matrix specified by
the $R$-matrix given by (\ref{R-matrix}) and the $K$-matrices given by (\ref{K-matrix-1}) and (\ref{K-matrix-2}) satisfies the crossing
symmetry (\ref{trans-crossing}).


\end{document}